%% file: BPH-11-022_temp.tex
\pdfoutput=1

\documentclass[11pt,twoside,a4paper,cmspaper,final,collab]{cms-tdr}
\def\svnVersion{253136}\def\cmsCernNo{2013-037}\def\cmsCernDate{\today}\def\cmsMessage{Published in the Journal of High Energy Physics as \href{http://dx.doi.org/10.1007/JHEP04(2012)084}{\doi{10.1007/JHEP04(2012)084.}}}

\begin{document}\cmsNoteHeader{BPH-11-022}

\hyphenation{had-ron-i-za-tion}
\hyphenation{cal-or-i-me-ter}
\hyphenation{de-vices}

\RCS$Revision: 183617 $
\RCS$HeadURL: svn+ssh://alverson@svn.cern.ch/reps/tdr2/papers/BPH-11-022/trunk/BPH-11-022.tex $
\RCS$Id: BPH-11-022.tex 183617 2013-04-29 19:11:16Z alverson $

\newcommand\deltar   {\ensuremath{\Delta R}}
\newcommand\nb   {\ensuremath{\,\mbox{nb}}}
\newcommand\ub   {\ensuremath{\,\mu\mbox{b}}}
\newcommand\ubNoSpace   {\ensuremath{\mu\mbox{b}}}
\newcommand\um   {\ensuremath{\,\mu\mbox{m}}}
\newcommand\tev  {\TeV}
\newcommand\gevNospace  {\GeVns}
\newcommand\gevc  {\GeVc}
\newcommand\gevcc  {\GeVcc}
\newcommand\mev  {\MeV}
\newcommand\mevc  {\MeVc}
\newcommand\mevcc  {\MeVcc}
\newcommand\Mb {\ensuremath{M_{\text{b}}}}
\providecommand\rd{\ensuremath{\mathrm{d}}}
\newcommand\dsdEt {\ensuremath{\rd\sigma/\rdE_\mathrm{T}^\mathrm{b}}}
\newcommand\dsdPt {\ensuremath{\rd\sigma/\rd\PT}}
\newcommand\dsdPtb {\ensuremath{\rd\sigma/\rdp_\mathrm{T}^\mathrm{b}}}
\newcommand\dsdeta {\ensuremath{\rd\sigma/\rd \abs{ \eta^\mathrm{b} }}}
\newcommand\dsdyb {\ensuremath{\rd\sigma/\rd \abs{ y^\mathrm{b}}}}
\newcommand\dsdy {\ensuremath{\rd\sigma/\rd\abs{y}}}
\newcommand\ptb {\ensuremath{p_\mathrm{T}^{\Pb}}}
\newcommand\Ptrel {\ensuremath{p_\mathrm{T}^{\text{rel}}}}
\newcommand\Etb {\ensuremath{E_\mathrm{T}^\mathrm{b}}}
\newcommand\Ptb {\ensuremath{p_\mathrm{T}^\mathrm{b}}}
\newcommand\etab {\ensuremath{\eta^\mathrm{b}}}
\newcommand\yb {\ensuremath{y^\mathrm{b}}}
\newcommand\y {\ensuremath{y}}
\newcommand\ayb {\ensuremath{\abs{ y^{\Pb}}}}
\newcommand\absetab {\ensuremath{\abs{\eta^\mathrm{b}}}}
\newcommand\absyb {\ensuremath{\abs{ y^\mathrm{b}}}}
\newcommand\abseta {\ensuremath{\abs{ \eta }}}
\newcommand\absy {\ensuremath{\abs{ y }}}
\newcommand\pl {\ensuremath{p_\mathrm{L}}}
\newcommand\T{\rule{0pt}{2.3ex}}
\newcommand\B{\rule[-1.0ex]{0pt}{0pt}}
\providecommand{\cPqb}{\ensuremath{\mathrm{b}}} 
\providecommand{\cPaqb}{\ensuremath{\overline{\mathrm{b}}}} 

\newcommand{\comment}[1]{}
\newcommand{\pb}{\ensuremath{\mathrm{pb}}}%
\newcommand{\pp}{\Pp\Pp\xspace}%
\newcommand{\rts}{\ensuremath{\sqrt{s}}}%
\newcommand{\invpb}{\mbox{$\textrm{pb}^{-1}$}}
\newcommand{\invnb}{\mbox{$\textrm{nb}^{-1}$}}
\newcommand{\hta}{\mbox{$ \eta$}}
\newcommand{\fh}{\mbox{$ \phi$}}
\newcommand{\Pt}{\pt}
\newcommand{\Et}{E_{T}}
\newcommand{\ppbb}{\Pp\Pp\rightarrow{\bbbar}X}
\def\ptmu{\ensuremath{{p^\mu_\mathrm{T}}}}
\def\etamu{\ensuremath{{\eta^\mu}}}

\def\vdef #1{\expandafter\def\csname #1\endcsname }
\vdef{n:result_vis}{113\pm 1 (\mathrm{stat}) \pm 14 (\mathrm{syst}) \pm 5 (\mathrm{lumi})}
\vdef{n:result}{2.25\pm0.01 (\mathrm{stat}) \pm 0.31 (\mathrm{syst}) \pm 0.09 (\mathrm{lumi})}
\vdef{n:nevents} {113\,561}
\vdef{n:lumi} {3}
\vdef{n:lumiError} {0.09}
\vdef{n:pythia}{3.27}
\vdef{n:mcnlo} {1.83^{+0.64}_{-0.42}({\mathrm{scale}}) \pm0.05({m_\text{b}}) \pm 0.08({\mathrm{pdf}})}
\vdef{n:ptmin} {9}

\cmsNoteHeader{BPH-11-022} 
\title{Inclusive b-jet production in pp collisions at $\sqrt{s}=7$~TeV}

\date{\today}
\def\vu   #1{\csname #1\endcsname}

\abstract{
  The inclusive b-jet production cross section
  in pp collisions at a center-of-mass energy of 7\TeV
  is measured using data
  collected by the CMS experiment at the LHC.
  The cross section is presented
  as a function of the jet transverse momentum
  in the range $18 < \pt< 200$\GeV for several rapidity intervals.
  The results are also given as the ratio of the b-jet production cross section
  to the inclusive jet production cross section.
  The measurement is performed with two different analyses, which differ in
  their trigger selection and b-jet identification: a jet analysis
  that selects events with a b jet using a sample corresponding to an integrated luminosity of 34\pbinv,
  and a muon analysis requiring a b jet with  a muon based on an integrated luminosity of 3\pbinv.
  In both approaches the b jets are identified by requiring a secondary vertex.
  The results from the two methods are in agreement with each other and with next-to-leading order
  calculations, as well as with predictions based on the \PYTHIA event generator.
}

\hypersetup{%
pdfauthor={CMS Collaboration},%
pdftitle={Inclusive b-jet production in pp collisions at sqrt(s)=7 TeV},%
pdfsubject={CMS},%
pdfkeywords={CMS, physics, jets, b-tagging}}

\maketitle 

\section{Introduction}
\label{sec:Intro}

The experimental measurement of the b-quark production cross section has been pursued
with interest at hadron colliders
because of discrepancies between theoretical predictions and experimental results,
e.g., at the Tevatron~\cite{Abachi:1994kj,Abbott:1999se,Abe:1993sj,Aaltonen:2009xn}
and at HERA~\cite{Adloff:1999nr,Aktas:2005zc,Chekanov:2008tx,Chekanov:2009kj}.
Substantial progress has been made in understanding the b-quark production
and fragmentation
processes, and the measurements are now in reasonable agreement with the
predictions in most regions of the phase space~\cite{Frixione:1997ma,Cacciari,Mangano,Frixione}.
Theoretical uncertainties are, however, sizable, and there is great interest in
verifying the results at the higher center-of-mass energies provided by the LHC.
Identification of b-quark jets by methods relying on the long b lifetime
is almost independent of the details of the fragmentation of a b quark
into a b hadron. Therefore, measuring the rate of b jets is a direct
measurement of the b-quark production rate, with a negligible systematic
uncertainty originating from fragmentation~\cite{Frixione:1996nh}.
In addition,
large logarithmic corrections due to hard collinear gluons are avoided when inclusive b jets are considered,
leading to more sensitive comparisons between experimental results and theoretical calculations.

First results on $\bbbar$ production in pp collisions at $\sqrt{s}=7\tev$ have been reported by the LHCb
Collaboration using semi-inclusive decays in the forward rapidity region~\cite{LHCb:2010},
and by the Compact Muon Solenoid (CMS) Collaboration~\cite{CMSExperiment}
using inclusive b $\rightarrow \mu$X decays~\cite{BPH-10-007} in the central rapidity region
and measuring the b-hadron production cross section as a function of the muon transverse momentum and
pseudorapidity. CMS has also measured the production cross sections of
fully reconstructed $\PBp$~\cite{BPH-10-004}, $\PBz$~\cite{BPH-10-005},
and $\PB_\cPqs$~\cite{BPH-10-013} mesons, as well as the angular correlations between
b and $\cPaqb$ hadrons,
based on secondary vertex reconstruction~\cite{BPH-10-010}.
The ATLAS Collaboration has measured the inclusive and dijet cross sections of b jets~\cite{atlasbjets}.

This paper presents CMS measurements of b-jet cross sections in several bins of jet rapidity
$y$
and transverse momentum $\Pt$. The b-jet cross section presented in this paper is defined as
the sum of the b and $\cPaqb$ jet contributions.
Two independent analyses are presented: a jet analysis, selecting events with a b jet,
and a muon analysis, requiring in addition a muon in the b jet. Despite the difference in
the corresponding integrated luminosity (34\pbinv and 3\pbinv, respectively), the precisions
of the two measurements are similar and dominated by systematic uncertainties, which differ
because of the use of different triggers and b-jet identification criteria.
Most of the analysis procedures are common in the two analyses, and the
differences are explained in the sections concerned.

The inclusive b-jet production cross section is also presented as the ratio to
the inclusive jet-production cross section measured by CMS in the same rapidity
intervals~\cite{QCD-10-011}.
The results are compared to theoretical predictions from next-to-leading order (NLO)
perturbative quantum chromodynamics (QCD)
calculations
and to predictions from the \PYTHIA event generator~\cite{Sjostrand:2006za}.

\section{CMS detector}
\label{sec:experiment}

The central feature of the CMS apparatus is a superconducting solenoid,
13~m in length and 6\unit{m} in diameter, which provides an axial magnetic
field of 3.8\unit{T}.  The bore of the solenoid is instrumented with various
particle detectors.  Charged particle trajectories are
measured with the silicon pixel and strip trackers,
covering $0 < \phi < 2\pi$  in azimuth and $|\eta |<$~2.5 in pseudorapidity,
where $\eta =-\ln[\tan{\theta/2}]$, with $\theta$ being the polar angle
of the track with respect to the counterclockwise beam direction.
The resolution is typically about 15\mum on the impact parameter and about
$1\%$ on the transverse momentum for charged particles with $\Pt < 40$\GeV.
A crystal electromagnetic calorimeter (ECAL) and a brass/scintillator hadron
calorimeter (HCAL) surround the tracking volume. The forward region is covered
by a an iron/quartz-fiber hadron calorimeter (HF).
The ECAL provides coverage in $|\eta| < 1.5$ in a cylindrical
barrel region and $1.5 < |\eta| < 3.0$ in two endcaps.
The ECAL has an energy resolution of better than 0.5\% for unconverted photons with transverse energies above $100~\GeV$.
The hadron calorimeters cover $|\eta| < 5.0$ with a jet energy resolution of about
$100\% / \sqrt{E}$, with the jet energy E in \GeV.
Muons are measured in gas-ionization detectors embedded in the steel return
yoke, covering $|\eta| < 2.4$.
A two-tier trigger system selects the most interesting pp collision events
for use in physics analyses.
A more detailed description of the CMS detector can be found
elsewhere~\cite{CMSExperiment}.

\section{Monte Carlo simulation}
\label{sec:simulation}

A detailed Monte Carlo (MC) simulation was performed for comparisons
with the data and to evaluate the selection efficiencies.
Simulated events were generated with \PYTHIA 6.422~\cite{Sjostrand:2006za} using tune Z2~\cite{Field:2010bc}
for the underlying event, a b-quark mass of $4.8\gev$, and the CTEQ6L1~\cite{Pumplin:2002vw}
proton parton distribution functions (PDF).
The generated events were processed through the full \GEANTfour~\cite{GEANT4} detector simulation,
trigger emulation, and event reconstruction chain.

The inclusive jet NLO theoretical prediction was
calculated with {\sc NLOJet++} \cite{Nagy:2001fj} using the CTEQ6.6M PDF set~\cite{Pumplin:2002vw}
and {\sc FastNLO} \cite{Kluge:2006xs} implementation.
The factorization and
renormalization scales were set to $\mu_F=\mu_R=\pt$. The inclusive b-jet
cross section prediction was calculated with \MCATNLO~\cite{Frixione:2002ik,Frixione:2003ei}
using the CTEQ6M PDF set and the nominal b-quark mass of 4.75\GeV. The
parton shower and hadronization were modeled using {\sc Herwig} 6.510~\cite{Marchesini:1991ch}.

The uncertainty on the predicted cross section was calculated independently by varying the
renormalization and factorization scales by factors of two, the b-quark mass by $\pm 0.25\gev$, and by
using the CTEQ6.6M instead of the CTEQ6M parton distribution functions~\cite{Pumplin:2002vw}.
\section{Event selection}
\label{sec:strategy}

The data used for this measurement were collected in 2010
and were required to pass the standard event quality criteria~\cite{QCD-10-011,BPH-10-007},
which reject data with anomalous or faulty behavior of the silicon tracker,
calorimeters, or muon chambers.
The total integrated luminosity amounts to $34\pbinv$ for the jet analysis
and to $3\pbinv$ for the muon analysis.

The inclusive jet data were collected using a combination of minimum bias
and single-jet triggers~\cite{CMSExperiment},
where each trigger covers a separate continuous \pt\ range (18--37, 37--56, 56--84,
84--114, 114--153, and 153--196\GeV, for trigger thresholds of 0, 6, 15, 30, 50, and 70\GeV in uncorrected $\pt$, respectively).
For each \pt\ bin, the trigger with the highest integrated luminosity is selected from those with $>$98\% efficiency at all rapidities.
For the muon analysis, the events are required to pass a trigger
selection~\cite{CMSExperiment} that accepts events with muons having $p^\mu_\mathrm{T}> 9\gev$ and $|\eta^\mu| < 2.4$.

Jets are reconstructed using a particle-flow algorithm~\cite{QCD-10-011}, which uses
the information from all CMS sub-detectors to reconstruct different types of particles produced
in the event. The basic objects of the particle-flow reconstruction are the tracks of charged particles
reconstructed in the central tracker, and energy deposits reconstructed in the calorimetry.
These objects are clustered with the anti-$k_\mathrm{T}$ algorithm~\cite{AntiKT,FastJet} using
the jet clustering distance parameter $R = 0.5$.
Tight jet identification criteria
~\cite{JME-10-011} are applied to
protect against poorly modeled sources of calorimeter noise. ß
The jet energies are corrected using estimates based on simulated events for the \pt\ dependence,
while corrections measured from data~\cite{JME-10-011} are applied for the absolute scale and the rapidity dependence.

The b jets are identified by finding the secondary decay vertex of the b hadrons~\cite{BTV-11-001}.
The secondary vertices from b- and c-hadron decays can be
distinguished by a selection on the relative distance from the primary vertex,
using the three-dimensional decay-length significance,
which is typically larger for b jets than for c, light-quark, and gluon jets.
In the jet analysis, a selection based on secondary vertices with at least three tracks containing signals from the
silicon pixel detector provide a clean signal against light-quark and gluon-jet
backgrounds. In the muon analysis, the minimum number of tracks to identify the secondary vertex is two,
in order to keep the b-tagging efficiency high
for semileptonic decays of b hadrons.

In the muon analysis, the offline selection requires at least one muon candidate in the \pt\ and $\eta$ ranges of the trigger selection
that fulfills a tight muon selection identical to that used in~\cite{BPH-10-007}.
The reconstructed muon is associated with the highest-\pt\ b-tagged jet within
a $\deltar = \sqrt{(\Delta \eta)^2 + (\Delta \phi)^2} < 0.3$ cone, where $\Delta \eta$ and $\Delta \phi$ refer
to the angular separation between the b-tagged jet and the muon.
If several
muons are associated with the b-tagged jet, the muon with the highest \pt\ is considered. According
to the simulation, the average efficiency of associating the muon with the b-tagged jet is $(76 \pm 2) \%$.
The probability of a random muon association with a jet is estimated to be less than 0.5\%.

The two b-jet cross-section measurement samples are collected with
different triggers and are essentially statistically independent.  The
effective trigger efficiency of the muon trigger is significantly higher, thereby
compensating for an order of magnitude smaller integrated
luminosity.  A total of 43\,046 jets pass the event and jet selection
for the jet analysis while in the muon analysis a total of 113\,561
events pass the event and jet selections, making the two analyses
comparable in terms of statistical power.
\section{Cross section measurement}
\label{sec:cross-section}

The production cross section for b jets is calculated as a double differential,
\begin{equation}
\frac{\rd^2\sigma}{\rd\pt \; \rd{y}} =
\frac{N_\text{tagged} \; f_\cPqb \; C_\text{unfold}}
    {\epsilon \; \Delta \pt \; \Delta y \; \mathcal{L}}\quad,
\end{equation}
where $N_\text{tagged}$ is the measured number of tagged jets per bin from the jet analysis
and the number of jets tagged with muons from the muon analysis,
$\Delta \pt$ and $\Delta y$ are the bin widths in \pt\ and $y$,
$f_\cPqb$ is the b-tagged sample purity,
$C_\text{unfold}$ is the unfolding correction,
and $\mathcal{L}$ is the integrated luminosity.
No distinction is made between b-quark jets and $\cPaqb$-quark jets,
so the cross section is the sum of b and $\cPaqb$ production.

The \pt\ spectra are normalized by the respective integrated luminosities of the individual jet triggers~\cite{QCD-10-011},
and then combined into a continuous jet \pt\
spectrum. Only one trigger is used for each \pt\ bin to simplify the analysis.
In the jet analysis, the reconstructed \pt\ spectra are unfolded using the ansatz
method~\cite{ansatz-one,ansatz-two}, with the jet \pt\ resolution obtained
with data-based methods from dijet data~\cite{JME-10-011}.
In the muon analysis, an unfolding (jet migration) correction derived from simulated events is applied to the selection efficiency as
the bin-by-bin ratio of the number of
generated b jets in a given \pt\ or rapidity bin
to the number of reconstructed b candidates in that bin.
In the simulation, the generated jets are constructed by clustering the stable particles produced during
the hadronization process including neutrinos in the muon analysis, but not in the jet analysis.
The two unfolding methods produce consistent results within the uncertainties of the jet $\Pt$ spectrum
and the jet $\Pt$-resolution modeling, which are negligible compared to the total systematic uncertainty.

The efficiency $\epsilon$  includes the trigger efficiency, event selection
efficiency, jet reconstruction and identification efficiency,
and the efficiency of tagging b jets. For the
muon analysis, the muon reconstruction efficiency is also included.

In the jet analysis, the efficiency is about 0.1\% to mistag light-quark and
gluon jets as b jets, and
the b-tagging efficiency is between
5\% at \pt\ $\approx$ 18\GeV and 56\% at \pt\ $\approx$ 100\GeV. The efficiency rises at higher
\pt\ as the average b-hadron
decay length increases.
To moderate the statistical fluctuations in the simulation, the
b-tagging efficiency in each rapidity bin is fitted to a functional parameterization versus $\pt$ accounting for
various effects such as the b-hadron proper time and the boost of secondary vertex decay products.
The fit
result is used in the analysis.
In the muon analysis, the average b-tagging efficiency is about 60\% in the barrel
region ($\absy <0.9$) and about 55\% for the endcap region
($1.2 < \absy <2.4$). It increases from 50\% to 75\%  for b-jet transverse momenta from 30 to 100\GeV.
The data/simulation scale factor for the b-tagging efficiency applied in the analysis is 0.95, with an uncertainty of 10\%~\cite{BTV-11-001}.

\begin{figure}[hbtp]
  \begin{center}
    \includegraphics[width=0.48\textwidth]{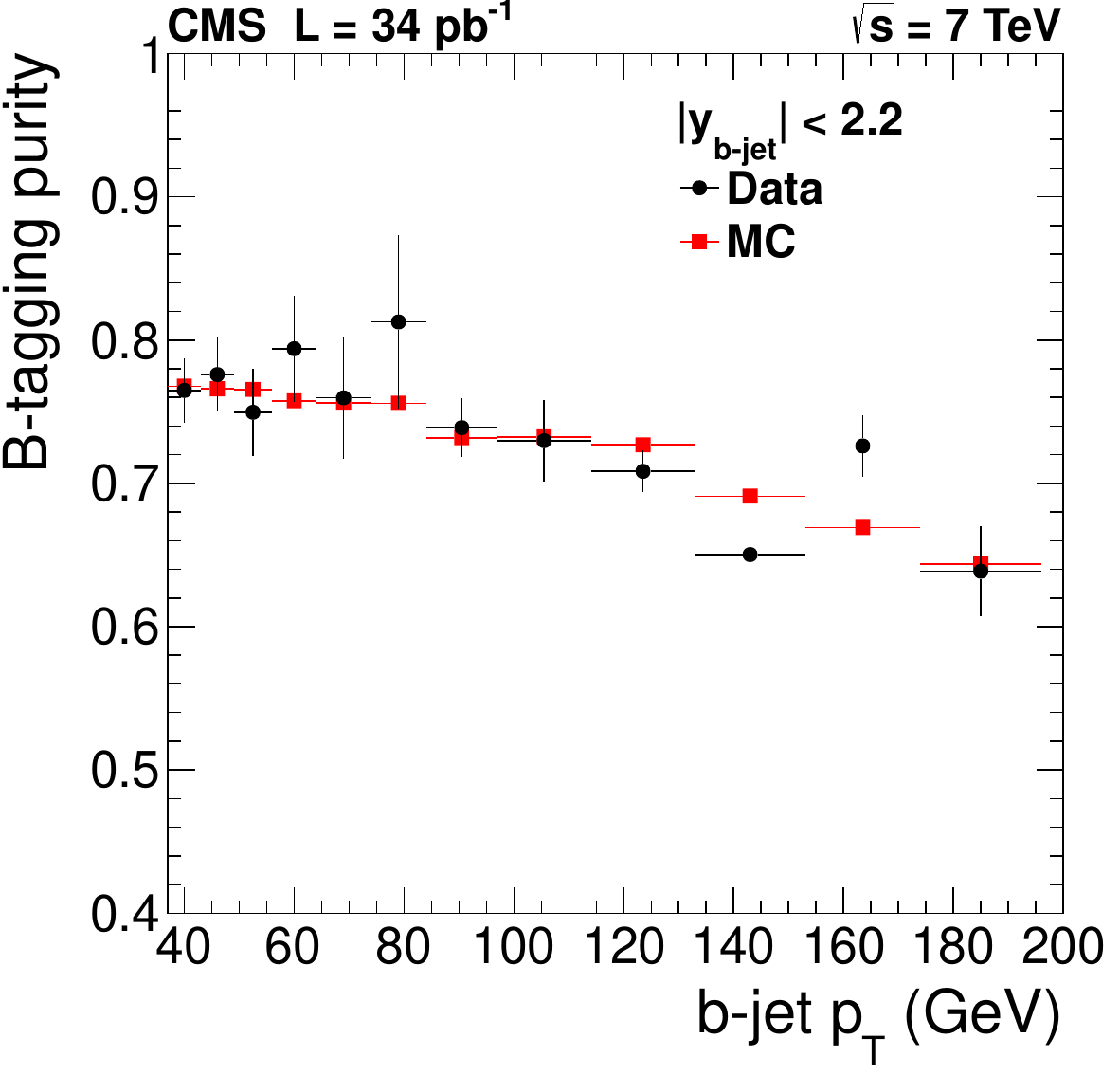}
    \includegraphics[width=0.46\textwidth,angle=90]{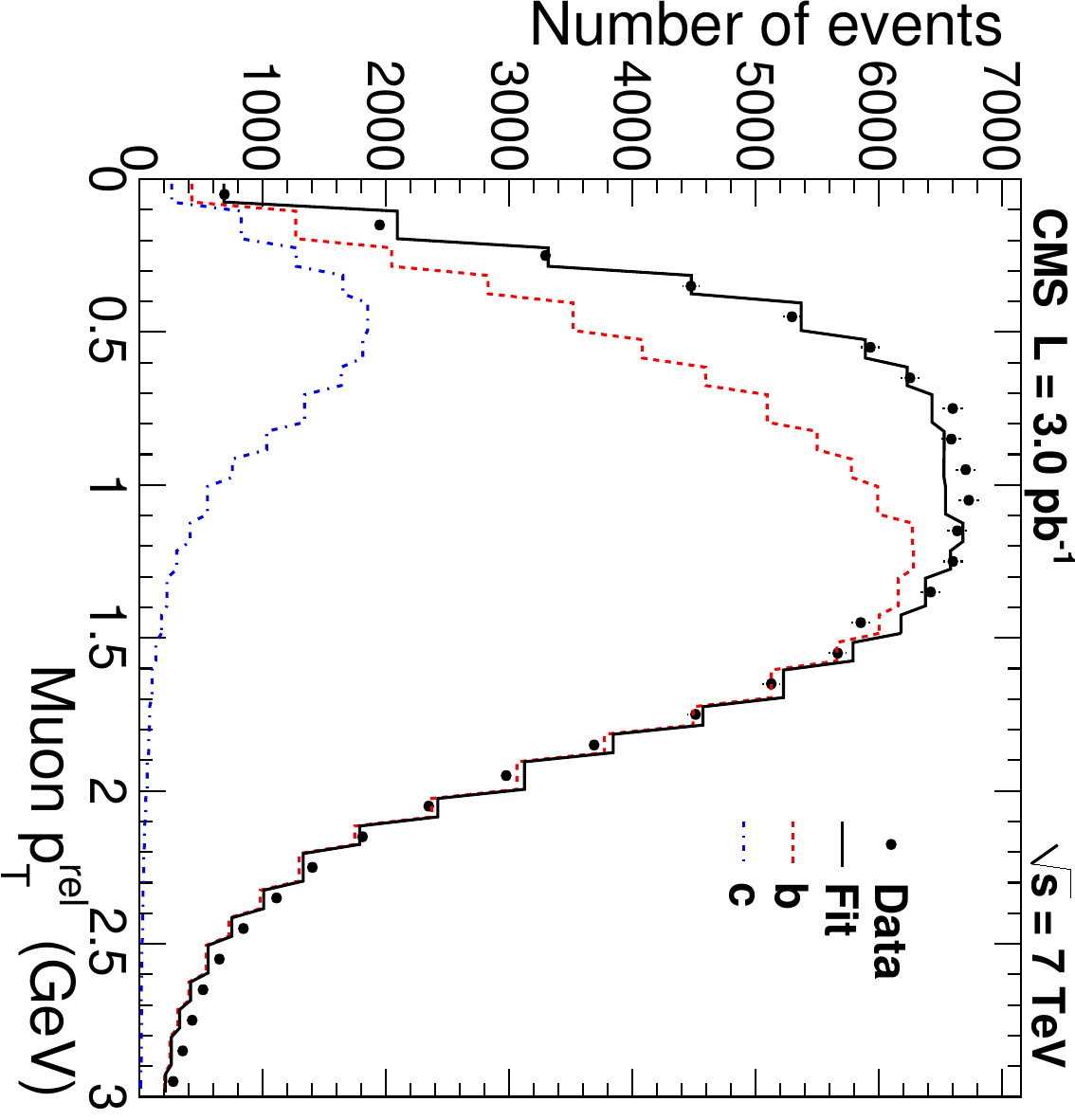}\quad%
    \caption{
    The b-tagged sample purity obtained using fits to the secondary vertex mass from data and simulated events
    as a function of the b-jet $\pt$ (left).
    The distribution of the muon  transverse momentum $\Ptrel$ with respect to the closest b-tagged jet in data for
    $\pt > 30\GeV$ and $\absy <2.4$, together with the maximum-likelihood fit (solid
    line) and its components (dashed lines) (right). The light-flavor (udsg) distribution is not visible in the figure
    since its contribution from the fit is consistent with zero.}
    \label{fig:fb2}
  \end{center}
\end{figure}

In the jet analysis, the distribution of the invariant mass of the tracks originating from the secondary vertex is fitted with probability density functions corresponding to vertex mass distributions for light-, charm-, and b-flavor jets taken from simulated events.
The relative normalizations for the combined light- and charm-flavor distribution and the b-flavor distribution are free parameters in the fit.
The resulting estimates of $f_\cPqb$ from data and simulated events
are shown in Fig.~\ref{fig:fb2} (left). 
The overall relative data/simulation scale factor
is consistent with unity within uncertainties.
Given the good agreement between data and simulation for $\pt > 37\GeV$,
the latter is used to predict the \pt\ and $y$ dependence of the purity, with no additional corrections, and to extrapolate
it to $\pt < 37\GeV$.

In the muon analysis, the b-tagged sample purity is obtained from a fit to the distribution of the
relative muon momentum \Ptrel\
with respect to the b-jet axis, which effectively
discriminates between b events and background. Figure~\ref{fig:fb2} (right) shows
the result of the \Ptrel\ fit, 
using the expected shapes from the simulated events for
the muons from b-hadron decays and background from charm quark and light-flavor jets.
The normalizations of the three contributions are
free parameters in the fit. A b fraction of $(86 \pm 5)\%$ is observed. The shapes obtained from the simulated events
provide a reasonable description of the data.
The  \Ptrel\ fit to the data gives a light-quark and gluon contribution to
the b-tagged jet sample of less than 3\% for all bins in $\pt$
and $\absy$.
This is confirmed in the simulated events where the
light-quark and gluon fraction of the b-tagged jets is estimated to
be less than $2\%$.

\section{Systematic uncertainties}
\label{sec:systematics}

The inclusive b-jet differential cross section can be affected by
uncertainties on the yield in each of the \pt\ bins and on the
measurement of the b-jet \pt\ itself, which determines the amount of smearing
between the neighboring bins and is corrected by unfolding.  The
leading uncertainties affecting the yields are due to the b-tagging
efficiency, the sample purity, and the integrated luminosity.  The
smearing of the \pt\ bin assignment is dominated by the jet energy
scale.
In the following, the systematic uncertainties common to the two analyses are discussed
first, and those specific to each analysis are then described separately.
All systematic uncertainties are summarized in Table~\ref{t:systematics}.

The uncertainty of the jet energy correction (JEC) is estimated using photon+jet events with the jet in the barrel
region, and dijet events where one jet is measured in the barrel region and the other in one of the endcaps~\cite{JME-10-011}.
These uncertainty estimates are further confirmed by indirect observations using comparisons of jet
substructure between data and MC simulations, the reconstruction of the $\Pgpz$ mass peak for the
ECAL energy scale,
and the measurement of the single-particle response in the calorimeters using isolated charged hadrons.
The uncertainty of the jet \pt\ resolution is estimated using
a comparison of dijet \pt\ balance between data and simulated events~\cite{JME-10-011}.

The cross-section measurement uses the b-tagging efficiency obtained from simulated events and
corrected by a scale factor measured in data.
Several methods based on muon-tagged jets~\cite{BTV-11-001} or $\ttbar$ events~\cite{BTV-11-003, TOP-10-003} are used
to measure the  b-tagging efficiency in data.
The ratio between the efficiencies measured from data and estimated from simulated events determines the scale factor
of $0.95 \pm 0.10$.

The difference between the inclusive-jet and the b-jet energy corrections is estimated
from MC fragmentation studies with \PYTHIA and \HERWIG to be 0.5--1.5\%, while studies based on data
find the inclusive jet scale uncertainty to be about 1.5--3.5\% for \pt\ $>30$\GeV and $|y|<2.2$.
Because of the lack of direct constraints from data on the relative b-jet energy scale, the b-jet and
the relative b-jet to inclusive JEC uncertainties are both taken to be the same as
the inclusive JEC uncertainty~\cite{JME-10-011}. Each 1\% uncertainty in the JEC translates into a 2--5\% uncertainty on the measured cross section
because of the steeply falling \pt\ spectrum.

Signals from the HF calorimeters are used to determine
the instantaneous luminosity
with a systematic uncertainty of 4\%~\cite{EWK-10-004}.

\begin{table}[!htb]
 \begin{center}
  \caption{Summary of the systematic uncertainties on the b-jet cross-section measurement, given in percent for the
    two analyses. The systematic uncertainties
    can vary depending on the b-tagged jet transverse momentum and
    rapidity, as indicated by the range in the table.}

  \begin{tabular}{l r r}
  \\
   \hline
   Source   &  Jet analysis & Muon analysis\\
   \hline
   Jet energy correction (JEC)   &  6--8  &  4--6   \\
   b-tagging efficiency          &  10--22  & 10 \\
   b sample purity      & 4--20   &   3--13   \\
   Luminosity                       &  4   &    4   \\

   Trigger efficiency                &  $<$ 1  &     3   \\
   Muon reconstruction efficiency    &  --  &     3   \\
   Selection      & $<$ 1   &  2--6   \\
   Muon-jet association    & --  &     2   \\
   b fragmentation     &  --  &  4   \\
   b $\rightarrow \mu$ branching fraction    & --  &     2.5   \\
   \hline
   Total                           & 13--24     &13--20   \\
   \hline
  \end{tabular}
  \label{t:systematics}
 \end{center}
\end{table}

\subsection{Systematic uncertainties specific to the jet analysis}

The b-tagged sample purity from the fit of the secondary vertex mass distribution and
the estimate from the simulated events are in agreement within 3--4\%.
The purity uncertainty is dominated by the uncertainty of the charm mistag rate across most of the kinematic range,
leading to a small uncertainty variation versus \pt\ and $y$. The light-flavor mistag rate is generally negligible
except at high \pt\ in the forward region,
where the yields are too low to perform a reliable fit.
This additional purity uncertainty is estimated by varying the light-quark and gluon mistag rate by $\pm$30\%.

The dominant source of uncertainty is the b-tagging efficiency.
In the ratio between the b-jet and the inclusive jet cross sections, the contribution from the luminosity uncertainty cancels,
and the impact of the jet energy resolution is negligible.
The contribution from the JEC in the ratio is not significantly reduced,
however, because the relative b-jet JEC is assumed to be uncorrelated with the inclusive
JEC. The JEC contributes 6--8\% to the total uncertainty.
The remaining systematic uncertainties from charm, light-quark, and gluon mistag rates
contribute 3--4\% to the b-tagged sample purity, except at high \pt\ and $y$, where a 30\% variation in the light-quark and gluon
mistag rate contributes up to 20\%.

The consistency of the simulation-based corrections for the b-tagging efficiency, the b-tagged sample purity,
the b-jet energy, and the inclusive jet energy scale, among others, is tested by running
the full analysis chain on reconstructed simulated events and comparing the results to
the particle-level \pt\ spectra. This closure test produces good agreement between the generated and
reconstructed \pt\ spectra to within 5\%.
This level of agreement is consistent with the statistical uncertainty of the simulation
and the systematic uncertainties of the parameterizations of the b-tagging efficiency and b-purity.

\subsection{Systematic uncertainties specific to the muon analysis}

The muon trigger efficiency is determined
from data using independent jet triggers. A systematic uncertainty of 3\% is assigned, which corresponds to the range
of differences between trigger efficiencies derived from data for muons from Z decays, muons in b-tagged events, and
muons with tight quality requirements.

The differences between the muon reconstruction efficiencies derived from data and simulated events is less than 2\%
in the barrel region and less
than 3\% in the endcap regions. A systematic uncertainty of 3\% is assigned for the muon reconstruction efficiency.

The efficiency for associating a muon with a b-tagged jet  agrees between data and simulation to within 2\%.

The uncertainties due to variations in the \Ptrel\ distributions between data and simulated events range from 3\% to 13\%.
This systematic uncertainty is estimated by varying the binning, including or not including the muon
in the definition of the jet direction, using different Monte Carlo simulation tunes,
and considering the overall difference between the data and fit results.
The largest contribution (up to 12\% for high-\pt\ b jets) is from the difference
between the signal fraction obtained by the \Ptrel\ fit and
by a fit to the secondary vertex mass distribution.

The uncertainty from the event selection is estimated from the variation of the muon selection cuts and the jet reconstruction,
and ranges from 2\% to  6\%.
The uncertainty of the b-quark fragmentation is determined by comparing the extrapolation factors to the total muon
transverse momentum range between \PYTHIA and \HERWIG~\cite{Corcella:2000bw}. It leads to a 4\% difference.
The branching fraction of b semileptonic decays into muons is known~\cite{Nakamura:2010zzi} to a precision of 2.5\%.
The signal fraction is also determined with an event selection based on calorimetric
jets~\cite{JME-10-011}. The measured cross sections are consistent within the systematic uncertainty.
The b fragmentation and b $\rightarrow \mu$ branching fraction uncertainties are taken into account only for
the b-jet cross section measurement extrapolated to cover the full \pt\ and $y$ range of the muons.
The total systematic uncertainty is 13\% at low jet \pt\ and increases to 20\% for high-\pt\ b jets.

\section{Results}
\label{sec:measurement}

\subsection{Jet analysis}
\begin{figure}[hbtp]
  \begin{center}
      \includegraphics[width=0.48\textwidth]{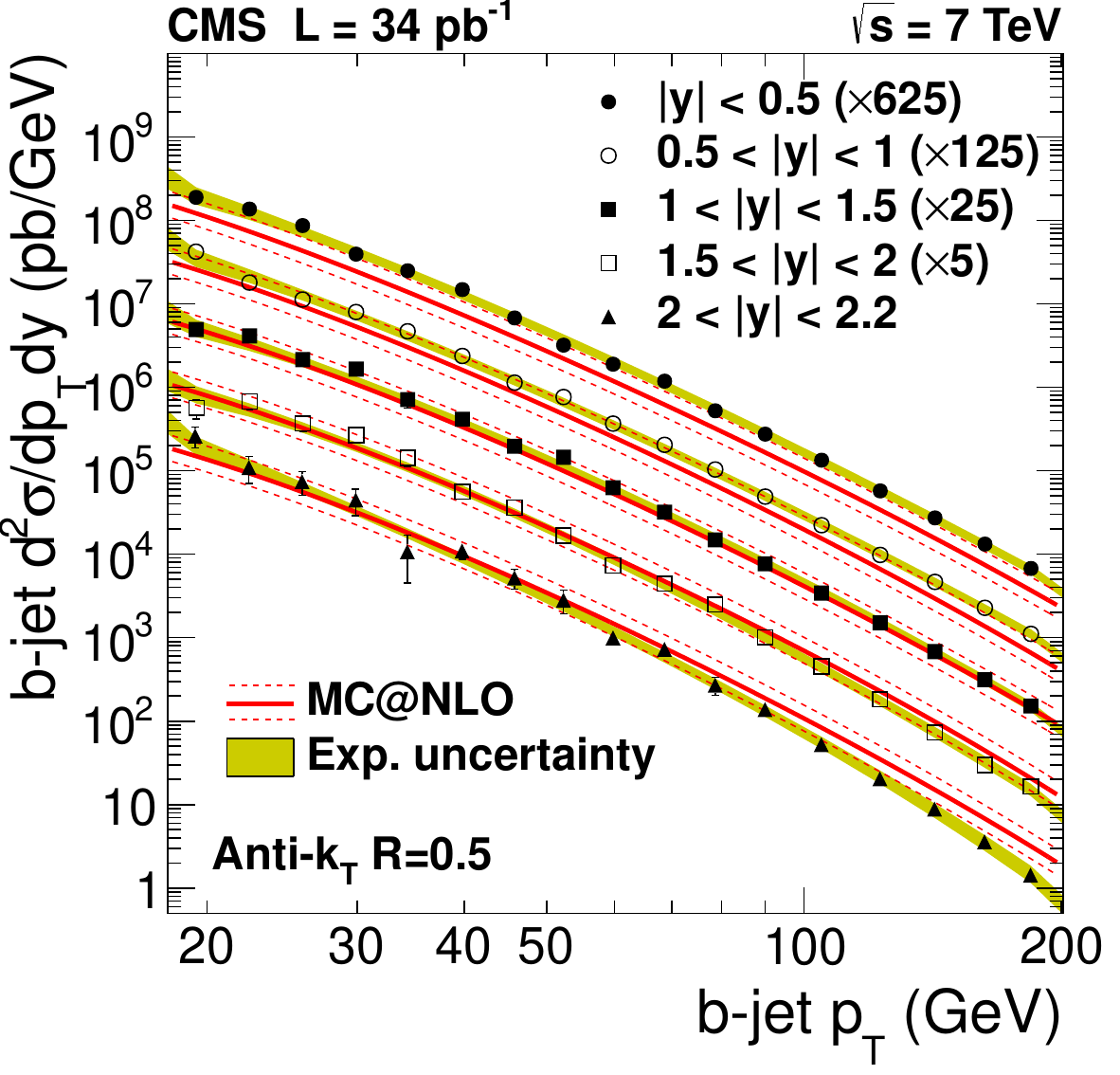}
      \includegraphics[width=0.48\textwidth]{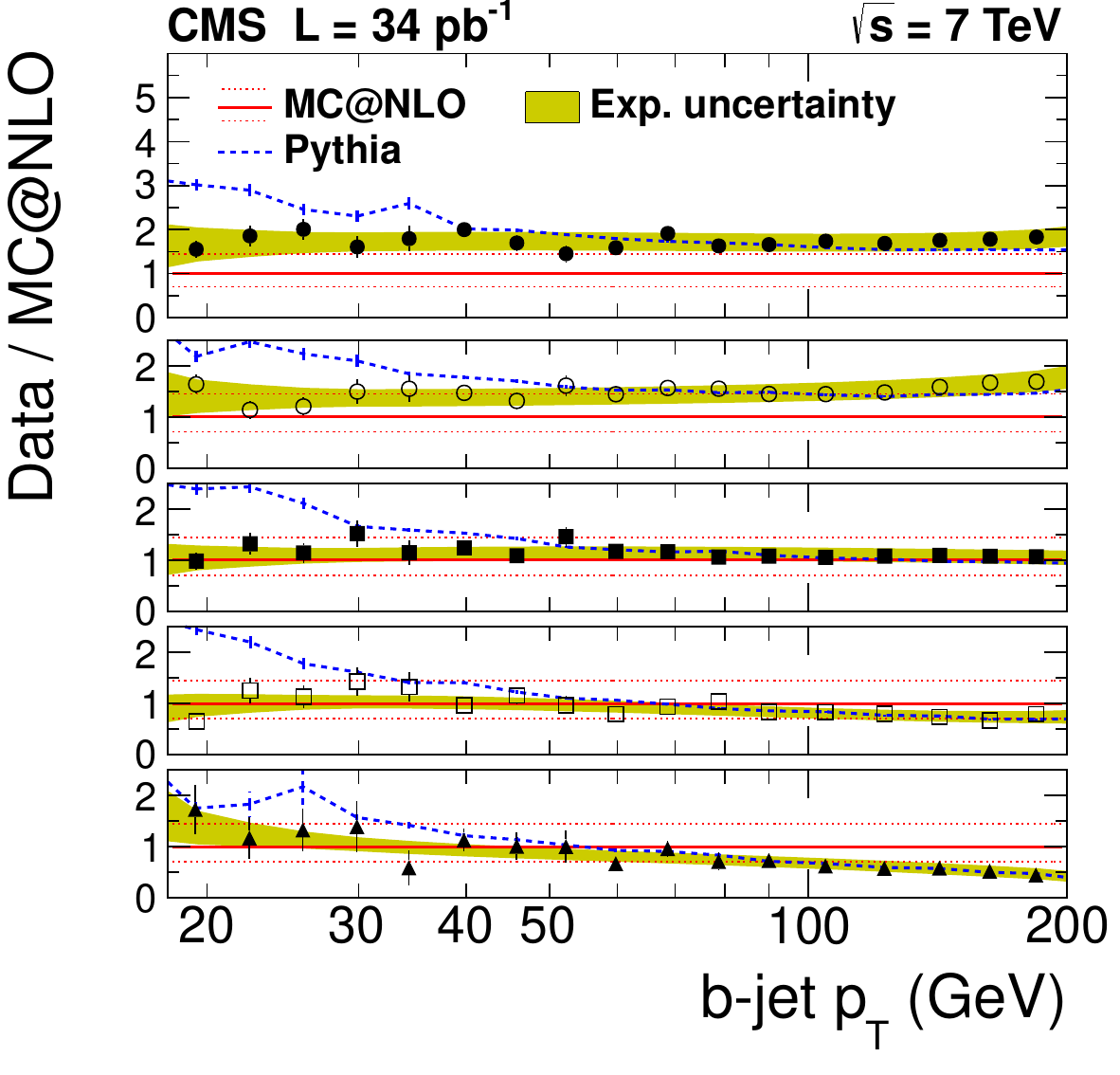}
    \caption{Measured b-jet cross section from the jet analysis, multiplied by the arbitrary factors shown in the figure
    for easier viewing, compared to the \MCATNLO calculation (left) and
    as a ratio to the \MCATNLO calculation
    (right).
The experimental systematic uncertainties are shown as a shaded band and the statistical uncertainties as error bars. The \MCATNLO uncertainty is shown as dotted lines. The \PYTHIA prediction is also shown in the right panel.}
    \label{fig:bcrossvs}
  \end{center}
\end{figure}

The measured b-jet cross section from the jet analysis is shown as a function of the jet \pt\
for different rapidity bins in Fig.~\ref{fig:bcrossvs} (left).
The values have been multiplied by the arbitrary factors given in the figure for easier viewing.
The cross section decreases by four orders of magnitude over the
\pt\ range 18--200\GeV. This behavior is well described by the theoretical
predictions from \MCATNLO, shown by the solid lines in the figure. Figure~\ref{fig:bcrossvs} (right) shows the ratio
between the measured cross section and the theoretical predictions.
The \MCATNLO values tend to be below the data in the central region ($|y|<1.0$) for
low \pt\ and above the data in the forward region ($|y|>2.0$) at large $\pt$.
The predictions from the \PYTHIA generator, in contrast, agree with the data at high $\pt$, but overestimate the cross section
significantly in the \pt\ region below 50\GeV, with the difference extending to higher \pt\
in the more forward region.

\begin{figure}[hbtp]
  \begin{center}
    \includegraphics[width=0.60\textwidth]{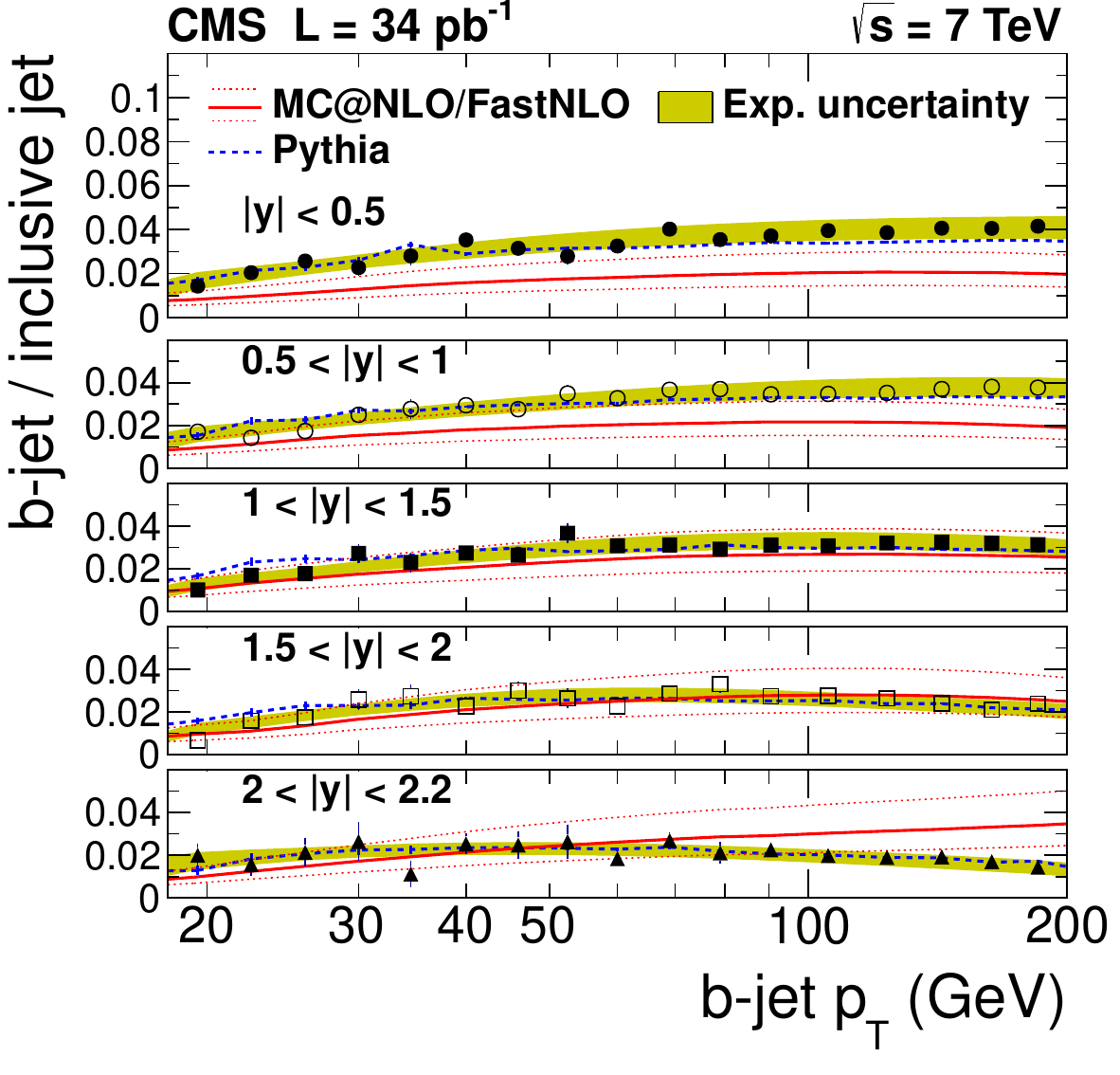}
    \caption{Ratio of the measured b-jet cross section from the jet analysis to the inclusive jet cross section~\cite{QCD-10-011},
as a function of the b-jet \pt\ (the jet \pt\ in the inclusive case). The predictions from NLO calculations (\MCATNLO/{\sc FastNLO}) and from \PYTHIA are also shown.}
    \label{fig:bcrossratio}
  \end{center}
\end{figure}

The ratio of the b-jet and the inclusive jet cross sections~\cite{QCD-10-011} is shown in
Fig.~\ref{fig:bcrossratio} as a function of $\pt$.
The ratio increases as a function of
\pt\ by up to a factor of 2, particularly in the central region.
The measurements are compared to the \MCATNLO prediction divided by
the {\sc FastNLO} prediction of the inclusive jet cross section~\cite{QCD-10-011}.
The non-perturbative corrections for the inclusive jet cross section prediction are
the average of {\sc  herwig6} \cite{Marchesini:1991ch} and \PYTHIA (tune D6T~\cite{Fano:2007zz}).
The data and NLO predictions agree within experimental and theoretical uncertainties.
Some difference between the NLO prediction and the data is observed in the central region, where the NLO values
are lower than the data, and at $\pt>100$\GeV and $|y|>2$, where the NLO prediction is higher than the data.
The \PYTHIA prediction for the ratio between the inclusive b-jet and inclusive-jet cross sections
is in agreement with the data across the full kinematic range of the measurement.

The total b-jet cross section is found by integrating the measured double-differential distributions over $|y|<2.2$ and
two different $\pt$
ranges: $18 < \pt < 200$\GeV and $32 < \pt < 200$\GeV. The values and the corresponding
\MCATNLO and \PYTHIA predictions are summarized in Table~\ref{t:xsection}.
\begin{table}[!htb]
 \begin{center}
  \caption{The b-jet cross sections (in \ub) measured from the jet and muon analyses.
The b-jet rapidity range is $\absy < 2.2$ and $\absy < 2.4$ for the jet and muon analyses, respectively.
The value for $\pt > 30\gev$ from the muon analysis is an extrapolated result.
For the data, the first uncertainty is statistical, the second is systematic,
and the third is associated with
the estimation of the integrated luminosity.
For the \MCATNLO prediction, the first uncertainty is from the variations in the QCD scale, the second from the b-quark mass, and the third from the
parton distribution functions.}

  \begin{tabular}{l p{2.2cm} c c c }
                    &          &  Data & \MCATNLO  &  \PYTHIA  \\
& & (\ub) & (\ub) & (\ub) \\ \hline
& & & & \\
Jet                    &  $\pt > 18\gev$ & $9.75 \pm 0.32 \pm 1.67 \pm 0.39$       & $7.3^{\,+2.9}_{\,-1.8} \pm 1.2 \pm 0.7$          & 15.3 \\
& & & & \\ \cline{2-5}
& & & & \\
                    &  $\pt > 32\gev$ & $1.73 \pm 0.07 \pm 0.20 \pm 0.07$       & $1.3 ^{\,+0.5}_{\,-0.3}\pm 0.2 \pm 0.1$          & 2.1 \\
& & & & \\ \hline
& & & & \\
Muon  & $\pt > 30\gev$ $\ptmu> 9\gev$ $|\etamu|<2.4$ & $0.113 \pm 0.001 \pm 0.014 \pm 0.005$ & $0.113^{\,+0.04}_{\,-0.023} \pm 0.003 \pm 0.005$ & 0.158 \\
& & & & \\ \cline{2-5}
& & & & \\
                    &  $\pt > 30\gev$ & $2.25 \pm 0.01 \pm 0.31 \pm 0.09$       & $1.83^{\,+0.64}_{\,-0.42}\pm 0.05 \pm 0.08$      & 3.27 \\
& & & & \\ \hline

  \end{tabular}
  \label{t:xsection}
 \end{center}
\end{table}

The \MCATNLO calculation predicts a
total $\ppbb$ cross section of 238\unit{$\mu$b}. The systematic uncertainty of this prediction comes
from varying the renormalization scale by factors of 0.5 and 2.0 ($+40$\%, $-25$\%), from variations
in the parameters of the  CTEQ PDF ($+10$\%, $-6$\%), and from the changing the b-quark mass
from 4.5 to 5.0\GeV ($+17$\%, $-14$\%). The total uncertainty on the theoretical calculation
is shown by the shaded bands in
Figs.~\ref{fig:bcrossvs} and \ref{fig:bcrossratio}.

\subsection{Muon analysis}

\begin{figure}[!htb]
  \centering
    \includegraphics[width=0.44\textwidth,angle=90]{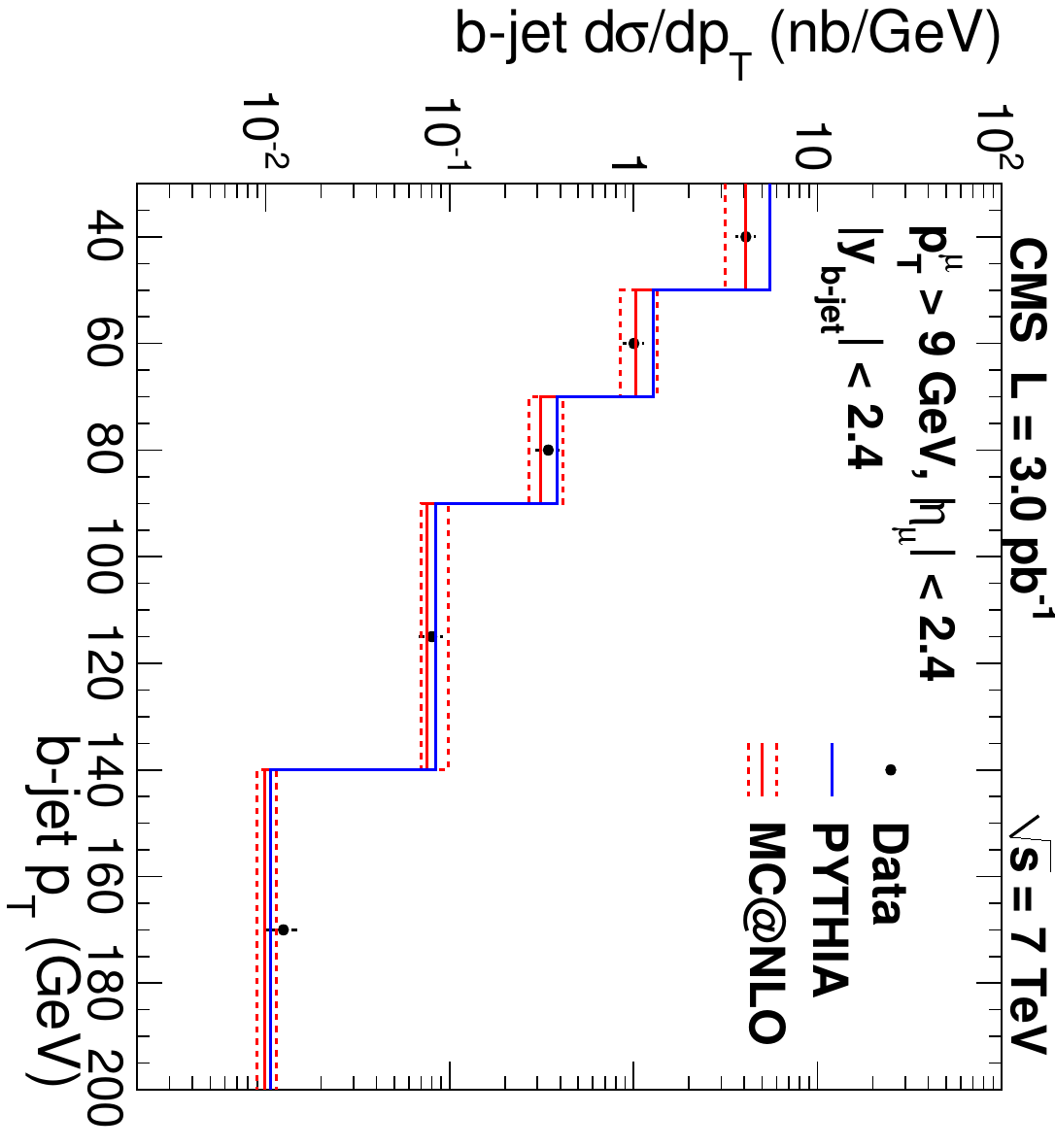}\quad%
    \includegraphics[width=0.44\textwidth,angle=90]{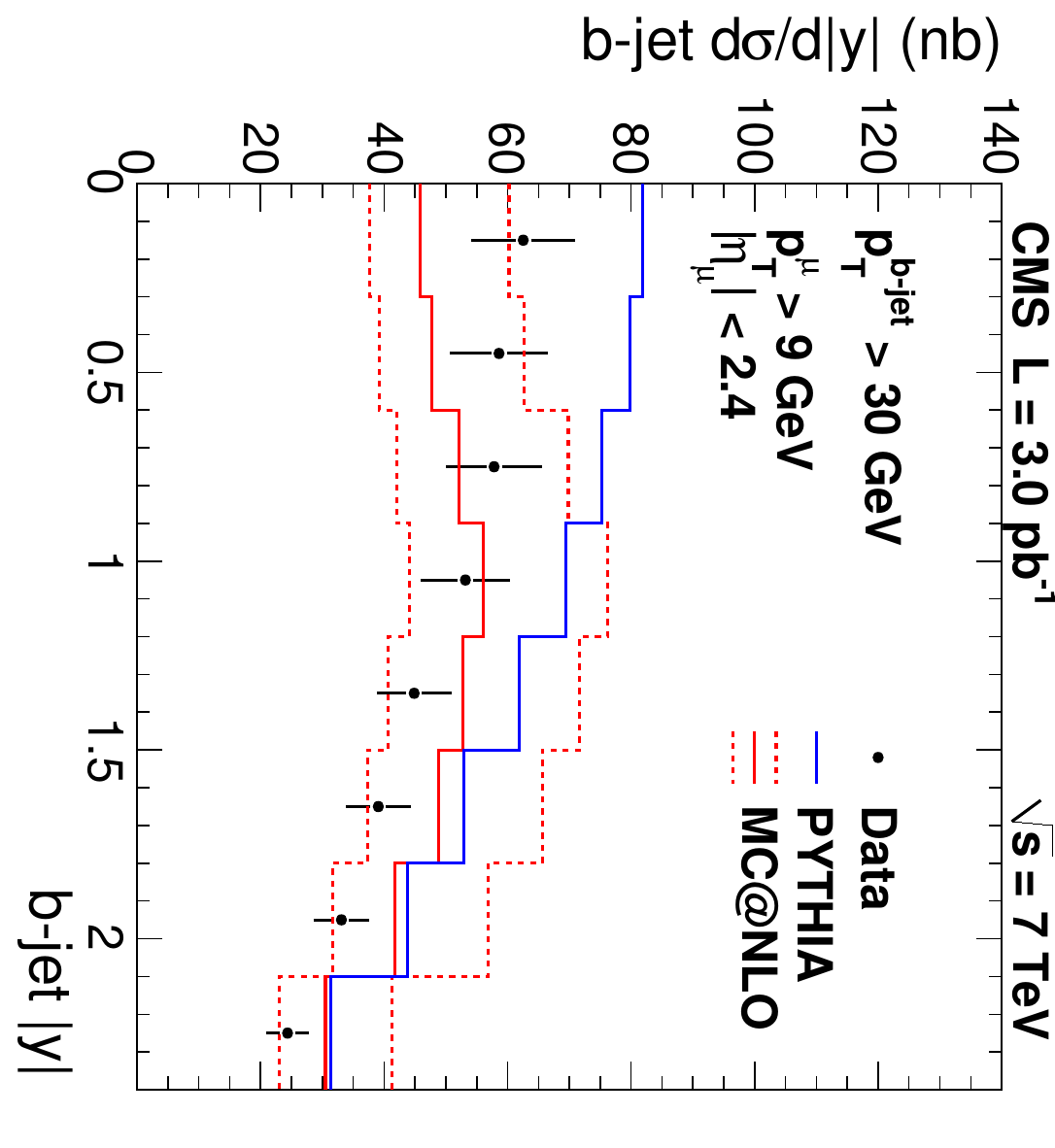}
    \caption{The differential b-jet cross section from the muon analysis as a function
of the b-jet \pt\ (left) and $\absy$ (right), with $\ptmu> 9\gev$ and
$|\etamu|<2.4$, and the predictions from \PYTHIA and \MCATNLO.
The error bars on the points correspond to the experimental statistical and systematic uncertainties added in quadrature.
The dashed lines represent the \MCATNLO uncertainty.
             \label{fig:inclb-bdsigmadpt_vis}}
\end{figure}

The measured differential cross sections for inclusive b-jet
production of b hadrons decaying into a muon with $\ptmu> 9\gev$ and
$|\etamu|<2.4$ are shown in Fig.~\ref{fig:inclb-bdsigmadpt_vis} as a function of the b-jet \pt\ (left) and
$|y|$ (right).
They are compared with the \MCATNLO and \PYTHIA predictions.
The dashed red lines illustrate the \MCATNLO theoretical uncertainty from variations in the QCD scale,
the b-quark mass, and the parton distribution functions.
A difference between the \PYTHIA prediction and the data is observed for b-jet $\pt < 70\gev$, where the \PYTHIA values are higher than the data.
The data are in agreement with the \PYTHIA prediction for the rapidity dependence of the cross section.
However, a significant difference in shape is observed between the data and the \MCATNLO predictions for the
rapidity dependence of the b-jet cross section.
A similar behavior had been observed in an inclusive b measurement with muons~\cite{BPH-10-007}.
The absolute normalization of the measured cross section is compatible with the NLO QCD predictions within the theoretical and experimental uncertainties.

The measured cross section for b jets with $\pt > 30\gev$, $\absy < 2.4$,
and the b hadrons decaying into muons in
the kinematic range $\ptmu> 9\gev$ and  $|\etamu|<2.4$,
is shown in Table~\ref{t:xsection}. The value is obtained by summing over all $\pt$ bins.

The measurements in the restricted muon kinematic range are extrapolated to cover
the full muon \pt\ and $y$ ranges using the \PYTHIA simulation, in order to obtain the b-jet cross section
limited only by the b-jet \pt\ and $y$.
The extrapolation also corrects for the branching fraction of b semileptonic decays into muons
and for the muon acceptance. The extrapolation factor changes from 5\% at low b-jet \pt\ to 20\%  at high \pt.
The \MCATNLO extrapolation factors are similar to those of  \PYTHIA at high b-jet \pt, while they
are about 20\% larger at low \pt.
The cross section measured in data and the corresponding \MCATNLO and \PYTHIA predictions are summarized in Table~\ref{t:xsection}.

\subsection{Comparison of results}

\begin{figure}[hbtp]
  \begin{center}
      \includegraphics[width=0.60\textwidth]{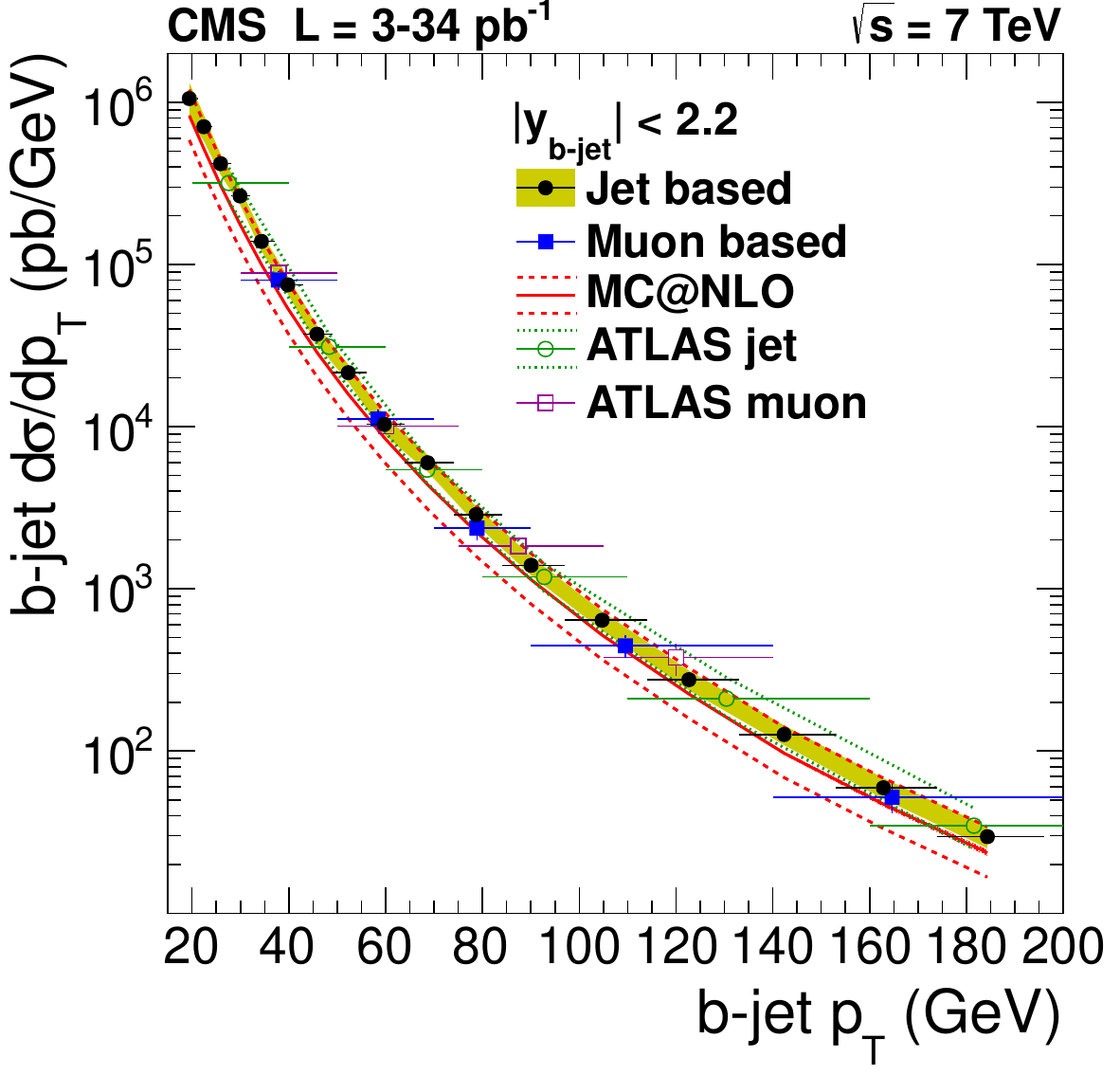}
    \caption{Measured b-jet cross sections in the jet and muon analyses as a function of the b-jet \pt,
    compared to the \MCATNLO calculation and to measurements from ATLAS~\cite{atlasbjets}.}
    \label{fig:comparison}
  \end{center}
\end{figure}

The measurements from the two analyses are compared in Fig.~\ref{fig:comparison} by adjusting the b-jet
cross section from the muon analysis 
to have the same visible phase space definition as the inclusive b-jet analysis, using \PYTHIA for the extrapolation.
The overall extrapolation factor is between 0.85 at $\pt = 30$\GeV and 0.82 at $\pt = 200$\GeV, and accounts for the reduction
in rapidity range from $|y|<2.4$ to $|y|<2.2$, exclusion of neutrinos from the particle jet definition,
and for counting all b-jets in the event. No additional uncertainty is assigned to the displayed cross sections
beyond the experimental uncertainties quoted in Table~\ref{t:systematics} and discussed in Section~\ref{sec:systematics}.
The closed circles in Fig.~\ref{fig:comparison} correspond to the measured inclusive b-jet
\pt\ spectrum, and
the closed squares show the b-jet \pt\ spectrum from the muon analysis, with
the yellow band representing the total experimental uncertainty.
Two sets of b-jet cross-section measurements from the ATLAS Collaboration~\cite{atlasbjets},
also found using a jet analysis and a muon analysis, are shown in the figure for comparison.
The CMS results are in good agreement with each other and with the ATLAS measurements to within their
respective uncertainties. The theoretical prediction from the NLO calculation~\cite{Frixione:2002ik,Frixione:2003ei} is displayed as the
solid line in the figure, with the dotted lines showing the systematic uncertainties.
The CMS results are consistent with the NLO predictions.

\section{Summary}
\label{sec:Conclusion}

The b-jet production cross section has been measured
in pp collisions at $\sqrt{s}=7$\TeV. The results were presented in several rapidity intervals as a function of the jet transverse momentum.
The results were also given as the ratio of the b-jet production cross section and the inclusive jet production cross section.
The results of two independent but compatible analyses were reported: a jet analysis selecting events with a b jet or a $\cPaqb$ jet, and
a muon analysis requiring in addition the presence of a muon,
based on integrated luminosities of 34\pbinv and  3\pbinv, respectively.

The measured values of the cross section were found to lie between the \MCATNLO and the \PYTHIA predictions.
The previous CMS measurements of $\PBp$~\cite{BPH-10-004}, $\PBz$~\cite{BPH-10-005}, and
$\PB_{\cPqs}$~\cite{BPH-10-013} production cross sections,
and an inclusive b-jet measurement with muons~\cite{BPH-10-007}, also gave values between these two predictions.
The measurement of the b-jet production cross section presented here will provide valuable input for testing
various theoretical models of b production and for further constraining their parameters.

\input{Acknowledgements}

\bibliography{auto_generated}   

\cleardoublepage \appendix\section{The CMS Collaboration \label{app:collab}}\begin{sloppypar}\hyphenpenalty=5000\widowpenalty=500\clubpenalty=5000\input{BPH-11-022-authorlist.tex}\end{sloppypar}
\end{document}

%% file: Acknowledgements.tex
\section*{Acknowledgements}
\label{sec:acknow}
\hyphenation{Bundes-ministerium Forschungs-gemeinschaft Forschungs-zentren} We wish to congratulate our colleagues in the CERN accelerator departments for the excellent performance of the LHC machine. We thank the technical and administrative staff at CERN and other CMS institutes. This work was supported by the Austrian Federal Ministry of Science and Research; the Belgium Fonds de la Recherche Scientifique, and Fonds voor Wetenschappelijk Onderzoek; the Brazilian Funding Agencies (CNPq, CAPES, FAPERJ, and FAPESP); the Bulgarian Ministry of Education and Science; CERN; the Chinese Academy of Sciences, Ministry of Science and Technology, and National Natural Science Foundation of China; the Colombian Funding Agency (COLCIENCIAS); the Croatian Ministry of Science, Education and Sport; the Research Promotion Foundation, Cyprus; the Ministry of Education and Research, Recurrent financing contract SF0690030s09 and European Regional Development Fund, Estonia; the Academy of Finland, Finnish Ministry of Education and Culture, and Helsinki Institute of Physics; the Institut National de Physique Nucl\'eaire et de Physique des Particules~/~CNRS, and Commissariat \`a l'\'Energie Atomique et aux \'Energies Alternatives~/~CEA, France; the Bundesministerium f\"ur Bildung und Forschung, Deutsche Forschungsgemeinschaft, and Helmholtz-Gemeinschaft Deutscher Forschungszentren, Germany; the General Secretariat for Research and Technology, Greece; the National Scientific Research Foundation, and National Office for Research and Technology, Hungary; the Department of Atomic Energy and the Department of Science and Technology, India; the Institute for Studies in Theoretical Physics and Mathematics, Iran; the Science Foundation, Ireland; the Istituto Nazionale di Fisica Nucleare, Italy; the Korean Ministry of Education, Science and Technology and the World Class University program of NRF, Korea; the Lithuanian Academy of Sciences; the Mexican Funding Agencies (CINVESTAV, CONACYT, SEP, and UASLP-FAI); the Ministry of Science and Innovation, New Zealand; the Pakistan Atomic Energy Commission; the Ministry of Science and Higher Education and the National Science Centre, Poland; the Funda\c{c}\~ao para a Ci\^encia e a Tecnologia, Portugal; JINR (Armenia, Belarus, Georgia, Ukraine, Uzbekistan); the Ministry of Education and Science of the Russian Federation, the Federal Agency of Atomic Energy of the Russian Federation, Russian Academy of Sciences, and the Russian Foundation for Basic Research; the Ministry of Science and Technological Development of Serbia; the Ministerio de Ciencia e Innovaci\'on, and Programa Consolider-Ingenio 2010, Spain; the Swiss Funding Agencies (ETH Board, ETH Zurich, PSI, SNF, UniZH, Canton Zurich, and SER); the National Science Council, Taipei; the Scientific and Technical Research Council of Turkey, and Turkish Atomic Energy Authority; the Science and Technology Facilities Council, UK; the US Department of Energy, and the US National Science Foundation.

Individuals have received support from the Marie-Curie programme and the European Research Council (European Union); the Leventis Foundation; the A. P. Sloan Foundation; the Alexander von Humboldt Foundation; the Belgian Federal Science Policy Office; the Fonds pour la Formation \`a la Recherche dans l'Industrie et dans l'Agriculture (FRIA-Belgium); the Agentschap voor Innovatie door Wetenschap en Technologie (IWT-Belgium); the Council of Science and Industrial Research, India; and the HOMING PLUS programme of Foundation for Polish Science, cofinanced from European Union, Regional Development Fund.

%% file: BPH-11-022-authorlist.tex
\textbf{Yerevan Physics Institute,  Yerevan,  Armenia}\\*[0pt]
S.~Chatrchyan, V.~Khachatryan, A.M.~Sirunyan, A.~Tumasyan
\vskip\cmsinstskip
\textbf{Institut f\"{u}r Hochenergiephysik der OeAW,  Wien,  Austria}\\*[0pt]
W.~Adam, T.~Bergauer, M.~Dragicevic, J.~Er\"{o}, C.~Fabjan, M.~Friedl, R.~Fr\"{u}hwirth, V.M.~Ghete, J.~Hammer\cmsAuthorMark{1}, M.~Hoch, N.~H\"{o}rmann, J.~Hrubec, M.~Jeitler, W.~Kiesenhofer, M.~Krammer, D.~Liko, I.~Mikulec, M.~Pernicka$^{\textrm{\dag}}$, B.~Rahbaran, C.~Rohringer, H.~Rohringer, R.~Sch\"{o}fbeck, J.~Strauss, A.~Taurok, F.~Teischinger, P.~Wagner, W.~Waltenberger, G.~Walzel, E.~Widl, C.-E.~Wulz
% ...
\vskip\cmsinstskip
\dag:~Deceased\\
1:~~Also at CERN, European Organization for Nuclear Research, Geneva, Switzerland\\
% ...
